\documentclass[acmlarge,nonacm]{acmart}

\usepackage{listings}

\usepackage{xcolor} 

\lstdefinelanguage{Solidity}{
    keywords=[1]{pragma, contract, function, returns, uint, if, else, for, while, mapping, address, require, emit},
    keywords=[2]{public, private, external, internal, view, pure, payable, constant, storage, memory},
    keywords=[3]{true, false, this, msg, block, tx},
    sensitive=true,
    comment=[l]{//},
    morecomment=[s]{/*}{*/},
    morestring=[b]{"}
}

\DeclareCaptionStyle{ruled}{labelfont=normalfont,labelsep=colon,strut=off}

\floatstyle{ruled}
\newfloat{listing}{tb}{lst}{}
\floatname{listing}{Listing}
\AtBeginDocument{%
  }

\begin{document}

\title{Implementation and Security Analysis of Cryptocurrencies Based on Ethereum}

\author{PENGFEI GAO}
\authornote{Both authors contributed equally to this research.}
\author{DECHAO KONG}
\authornotemark[1]
\affiliation{%
  \institution{Hainan University}
  \city{Haikou}
  \country{China}
}
\email{3037317181@gmail.com}

\author{XIAOQI LI}
\affiliation{%
  \institution{Hainan University}
  \city{Haikou}
  \country{China}}
\email{csxqli@ieee.org}


\begin{abstract}
Blockchain technology has set off a wave of decentralization in the world since its birth. The trust system constructed by blockchain technology based on cryptography algorithm and computing power provides a practical and powerful solution to solve the trust problem in human society. In order to make more convenient use of the characteristics of blockchain and build applications on it, smart contracts appear. By defining some trigger automatic execution contracts, the application space of blockchain is expanded and the foundation for the rapid development of blockchain is laid. This is blockchain 2.0. However, the programmability of smart contracts also introduces vulnerabilities. In order to cope with the insufficient security guarantee of high-value application networks running on blockchain 2.0 and smart contracts, this article will be represented by Ethereum to introduce the technical details of understanding blockchain 2.0 and the operation principle of contract virtual machines, and explain how cryptocurrencies based on blockchain 2.0 are constructed and operated. The common security problems and solutions are also discussed. Based on relevant research and on-chain practice, this paper provides a complete and comprehensive perspective to understanding cryptocurrency technology based on blockchain 2.0 and provides a reference for building more secure cryptocurrency contracts.
\end{abstract}




\maketitle

\section{Introduction}

Blockchain is a specialized form of distributed data storage that was first introduced as the underlying technology of Bitcoin in the paper Bitcoin: A Peer-to-Peer Electronic Cash System, published in 2008 by an individual or group under the pseudonym Satoshi Nakamoto during the subprime mortgage crisis. This technology pioneered a novel solution to the trust problem in distributed ledger storage through the combination of hash chaining and the proof-of-work mechanism. Due to its characteristics of data transparency, decentralization, and immutability, blockchain has been widely adopted in decentralized digital currency issuance and payment systems. The decentralization of digital currencies initiated and exemplified by Bitcoin is referred to as Blockchain 1.0\cite{mukherjee2021blockchain}. To enable blockchain to support more complex applications, programmable smart contracts were first introduced on top of the blockchain ledger structure, allowing Turing-complete programs to run on-chain. This advancement facilitated the development and execution of more sophisticated applications directly on blockchain platforms, significantly expanding its application scope and laying the foundation for its rapid evolution—this stage is known as Blockchain 2.0\cite{aggarwal2021blockchain}.\par

The most representative example of Blockchain 2.0 is Ethereum, which introduced a novel mechanism for token crowdfunding, commonly known as Initial Coin Offerings (ICOs). Developers can effortlessly create their own tokens on Ethereum via smart contracts—so-called programmable tokens—where the on-chain services provided by contracts serve as a fundamental value anchor for these tokens. Under this paradigm, the integration of smart contracts with digital assets has fostered a dynamic, decentralized, and multi-layered financial ecosystem known as Decentralized Finance (DeFi). As illustrated in Figure 1, taking the Ethereum ecosystem as an example, this ecosystem is structured around Ethereum’s native currency, ETH, as Layer 0. With Ethereum 2.0 supporting ETH staking, a bond market emerges, where interest rates regulate the flow of ETH throughout the ecosystem. Built upon this market, Layer 1 establishes a stability layer, ensuring stable value. MakerDAO, for instance, employs Collateralized Debt Position (CDP) contracts to lock ETH and generate DAI, a stablecoin pegged to the U.S. dollar\cite{li2024cobra}. Consequently, Layer 1 also functions as a capital formation layer, enabling individual users to participate in token minting. Moving up, Layer 2 represents the capital utility layer, where DAI-based lending mechanisms regulate borrowing costs through interest rate balancing. These tokens subsequently flow into the application layer, providing liquidity for various decentralized applications. At this level, atomic financial services such as token exchanges (Uniswap), prediction markets (Augur), and derivatives trading platforms (dYdX) operate seamlessly. Ultimately, the ecosystem culminates in a user aggregation layer, facilitating cross-chain transactions, credit card integrations, real estate exchanges, and other financial services.Even within the realm of decentralized finance alone, Blockchain 2.0 has demonstrated immense vitality, giving rise to a sophisticated value-driven internet. At the heart of this thriving value network lies smart contracts, which serve as the foundational infrastructure. Programmable tokens have unlocked boundless possibilities for cryptocurrencies, cementing blockchain's role as a transformative force in the digital economy.\par

\begin{figure*}[htbp]
     \centering
    \includegraphics[height=10cm,width=16cm]{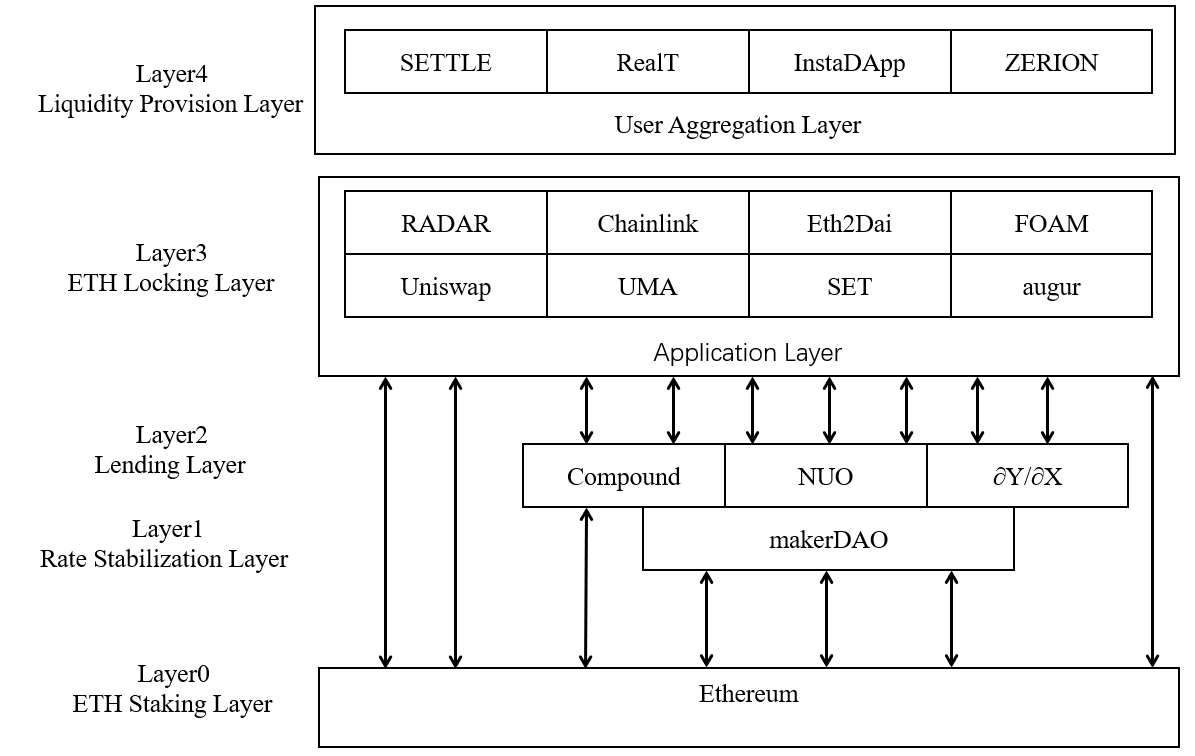}
    \caption{Layered Architecture of the Ethereum Ecosystem.}
    \label{fig:1}
\end{figure*}

As the Value Internet continues to expand, an increasing amount of capital has flowed into tokenized assets, leading to an exponential surge in token valuations over recent years. Meanwhile, the security concerns associated with smart contracts have become increasingly critical. On-chain token contracts are immutable, meaning that once deployed, all associated transactions are irrevocably determined. While this immutability ensures fairness and transparency, it also introduces significant risks\cite{curry2025limitations}. If vulnerabilities exist in the smart contract underpinning a token, any security breach could result in substantial digital asset losses, with no way to reverse or amend the deployed contract\cite{li2024scla}. In June 2016, the large-scale The DAO project, which raised \$150 million through an ICO within a month, was found to have a reentrancy vulnerability in its smart contract\cite{ma2025understanding}. Exploiting this flaw, hackers drained \$60 million worth of Ether, causing a sharp drop in Ethereum’s market price and ultimately leading to a hard fork of the Ethereum blockchain. In August 2016, the major exchange Bitfinex was attacked, resulting in the theft of 119,756 Bitcoin, valued at approximately \$65 million at the time.In July 2017, the widely used Parity Ethereum wallet was compromised, leading to the theft of 150,000 Ether, worth \$30 million. Later that year, in November, another vulnerability in the Parity wallet resulted in 513,701 Ether being permanently locked. In April 2018, the BEC and SMT token contracts fell victim to an integer overflow attack, allowing hackers to mint and dump massive amounts of tokens, effectively reducing their value to near zero.In April 2020, the Lendf.Me lending protocol was exploited due to reentrancy issues and security flaws in its unique token composition, leading to a total depletion of assets from the contract, with losses amounting to \$25 million.\par

In recent years, significant progress has been made in blockchain and smart contract security research both domestically and internationally\cite{dika2018security, kushwaha2022ethereum}. Security-compliant audits and formal verifications have substantially reduced the occurrence of major financial incidents on blockchain networks. However, security breaches leading to economic losses remain frequent. Analyzing various attacks on smart contracts reveals that while the underlying blockchain technology is inherently robust and rarely encounters critical failures, vulnerabilities are prevalent at the smart contract layer. These issues primarily stem from two factors: inherent logical flaws introduced by the programmability of smart contracts and security vulnerabilities arising from interactions between smart contracts and the contract virtual machine. The application of blockchain in currency-related use cases necessitates heightened attention to potential risks. Any oversight in the development of smart contracts can result in the deployment of insecure applications onto the blockchain, where their immutable nature makes rectification nearly impossible, leading to irreversible financial losses\cite{duy2025vulnsense, liang2025vulseye}. Therefore, research on the security of smart contracts in Blockchain 2.0 is of critical significance. As shown in Figure 2, the second section of this paper will begin by discussing the foundational mechanisms that support blockchain, progressively introducing the various technologies within the Blockchain 2.0 system that incorporates smart contracts. In the third section, the Ethereum smart contract system will be presented as a typical example of Blockchain 2.0, along with its model. The fourth section will explore the implementation of cryptocurrencies in the form of smart contracts, and finally, in the fifth section, we will replicate significant security incidents that occurred on the Ethereum blockchain by writing a simple token contract, analyzing common security issues, and discussing feasible solutions and mitigation strategies.\par

\begin{figure*}[htbp]
     \centering
    \includegraphics[height=9cm,width=14cm]{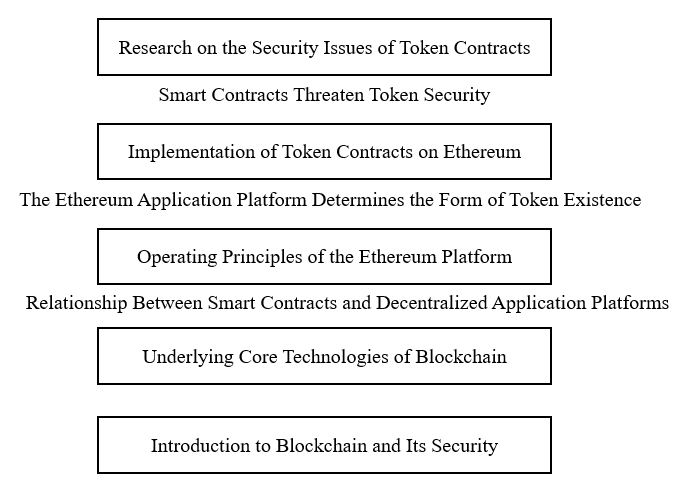}
    \caption{Structural Framework of Token Contract Security Research.}
    \label{fig:2}
\end{figure*}

\section{Background}
\subsection{Blockchain}Blockchain technology is a new distributed infrastructure and computing paradigm that uses a block-based data structure to verify and store data, employs a consensus algorithm from distributed nodes to update data, ensures the security of data transmission and access through cryptography, and utilizes smart contracts (automated script code) to program and manipulate data.\par

Blockchain establishes a decentralized model under zero trust, making it the core of encrypted digital currencies\cite{liu2022blockchain}. The main components of blockchain include blocks, chains, and the operations stored within them, namely transactions.\par

\begin{itemize}
\item {\texttt{Block}}: A block records transactions and states within a specific period, serving as the fundamental storage unit of the blockchain and a collection of transactions that have been completed.
\item {\texttt{Chain}}: A chain structure that links blocks in chronological order using hashes.
\item {\texttt{Transaction}}: All operations in the blockchain network are treated as transactions, also known as operations, through which records are generated on the blocks and the corresponding states are altered.
\end{itemize}

\subsection{Typical blockchain hierarchical structure}

As shown in Figure 3, from the perspective of the blockchain hierarchical structure, the blockchain is composed of the data layer, network layer, consensus layer, incentive layer, contract layer, and application layer, from bottom to top\cite{al2022hierarchical}. At the data level, the blockchain structure with chain-based storage and Merkle trees stores data in parallel, ensuring the integrity of the content through technologies such as hashing, digital signatures, and asymmetric encryption. Transactions initiated by clients are broadcasted in the P2P network after verification and temporarily stored as unconfirmed transactions on all nodes. Every period, machines across the network package all transactions within that time slice into blocks, and consensus is reached at the consensus layer, ensuring consistency of the data on the blockchain across all nodes in the network\cite{yu2020blockchain}. The typical consensus mechanism in the blockchain 1.0 era was Proof of Work, and once the block achieved consensus, it possessed the characteristic of immutability\cite{zhang2020overview}. Later, Ethereum introduced the Proof of Stake (POS) mechanism. The operation of the blockchain relies on miners, and at the incentive layer, virtual currency issuance and redistribution are realized. Both running contracts and transactions require paying miners with Ether as transaction fees. On this basis, blockchain 2.0 introduces smart contracts, which invoke contracts through data embedded in transactions. When packaging blocks, the defined virtual machine executes the contract scripts, enabling complex programmable functions. Finally, various user applications are realized through the interfaces provided by the smart contracts\cite{wang2019blockchain}.\par

\begin{figure*}[htbp]
     \centering
    \includegraphics[height=11cm,width=14cm]{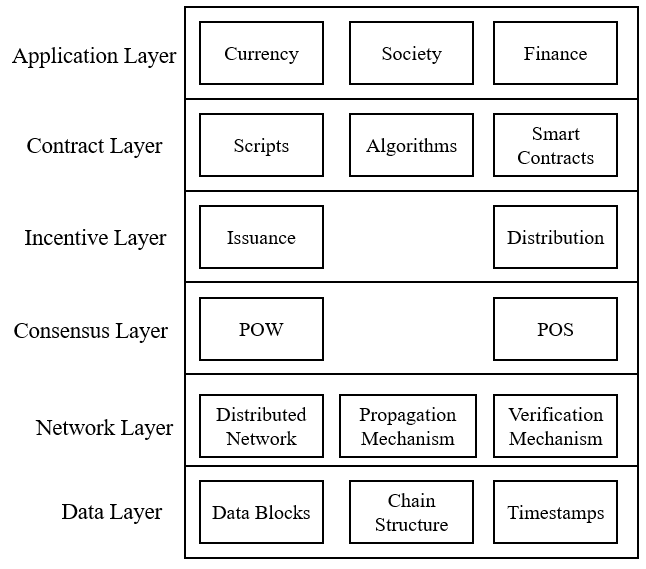}
    \caption{Layered Technical Stack of Blockchain Systems.}
    \label{fig:3}
\end{figure*}

\subsection{Consensus Mechanism}
To address the trust issue in distributed networks, an effective consensus mechanism is required to balance system efficiency and usability.\par

The consensus mechanism used by early virtual currency systems such as Bitcoin and the early stages of Ethereum is Proof of Work (PoW), which ensures data consistency and consensus through computational power competition among network nodes\cite{ferdous2021survey}. All nodes attempt to find a random number (Nonce) that makes the hash of the current block smaller than a certain value or starts with a specific number of zeros. When a node discovers this random number, it gains the right to record the block and is rewarded as a miner. The essence of "mining" in blockchain is the computational brute force to break the hash. The Proof of Work mechanism can secure the blockchain, ensuring that the accounting rights are random and providing effective protection for the secure operation of blockchain systems like Bitcoin, as long as more than 51\% of the computational power is not controlled maliciously. Compared to Proof of Work, Proof of Stake significantly improves energy efficiency by eliminating the need for large-scale ASIC computational power competition\cite{sedlmeir2020energy, budish2018economic, andoni2019blockchain}. At the same time, it greatly increases the cost of attacks and lowers the entry barriers for participating in blockchain maintenance, thus enabling more individuals to get involved and providing stronger decentralization.\par

\section{The operating principles of the Ethereum platform}
This paper takes Ethereum as a typical representative of Blockchain 2.0 and studies the security issues of smart contracts under Blockchain 2.0. Therefore, it first introduces the technical details of the Ethereum platform.\par

\subsection{Ethereum Application Technology}
\subsubsection{Ethereum Virtual Machine}
The Ethereum Virtual Machine serves as the execution environment for smart contracts\cite{hirai2017defining}. All smart contracts and state modifications on the Ethereum blockchain are conducted through transactions, and every transaction on the Ethereum network is executed by the EVM. The EVM introduces an abstraction layer above Ethereum nodes, enabling Turing completeness through the execution of specific operations defined by 140 opcodes\cite{hildenbrandt2017kevm, fekete2023toward}. This capability allows the EVM to support the deployment and execution of various smart contracts with diverse functionalities [16]. Essentially, the Ethereum Virtual Machine is a stack-based machine, with its primary function being the execution of smart contracts. It operates with a 256-byte word size, a stack depth of 1024, and is designed with simplicity, determinism, space efficiency, blockchain-oriented functionality, security assurances, and optimization in mind. Data within the EVM can be stored in three distinct locations: the stack, temporary storage, and persistent storage\cite{albert2024superstack, mazumdar2019survey}. The EVM executes operations by interpreting opcodes within smart contracts, manipulating on-chain and transaction data to produce a deterministic and unique outcome.\par

\subsubsection{Smart Contract}
A computer program capable of automatically enforcing contractual terms is defined as a smart contract. Although the concept of smart contracts emerged almost simultaneously with the internet, there was no perfect and reliable technological solution to ensure the security and trustworthiness of contract execution until the advent of blockchain technology\cite{khan2021blockchain, christidis2016blockchains}. In the context of Ethereum, a smart contract is an executable program that runs on the Ethereum blockchain. These contracts are stored on-chain and assigned a unique address. Their execution is triggered by transactions sent to this address, incurring computational costs and modifying the blockchain state. Smart contracts also serve as a public interface for interactions between users and decentralized applications. Once deployed, a smart contract remains persistently available, is difficult to modify, and cannot be revoked.\par

\subsubsection{Ethereum Nodes}
For an application to interact with the Ethereum blockchain, it must connect to an Ethereum node, which serves as the gateway to the entire Ethereum network. An Ethereum node is a computer running an Ethereum client—an implementation of the Ethereum protocol capable of validating all transactions within each block, thereby ensuring network security and data accuracy. Ethereum nodes collectively maintain the state of the blockchain and achieve consensus on state changes through the underlying consensus algorithm\cite{wang2019survey}. By facilitating communication between applications and the blockchain, Ethereum nodes play a crucial role in maintaining the integrity and functionality of the Ethereum ecosystem.\par

\subsubsection{Ethereum Client API}
Applications connect to and communicate with the Ethereum blockchain through API libraries developed and maintained by the Ethereum open-source community. These APIs abstract much of the complexity associated with direct interaction with Ethereum nodes, significantly reducing the technical burden on developers. Additionally, these libraries provide convenient functions that allow developers to spend less time dealing with the intricacies of Ethereum clients and instead focus on implementing the business logic of their applications.\par

\subsubsection{End-User Applications}
At the top of the stack are user-facing applications, which primarily include two common types: web applications and mobile applications. Due to well-designed encapsulation and development, users often do not need to be aware that the applications they are using are built on blockchain technology.\par

\subsection{Ethereum Blockchain Model}
Compared to the Bitcoin system, Ethereum, despite having a similar cryptocurrency (ETH) that follows nearly identical intuitive rules, offers more powerful functionality through smart contracts. Rather than merely serving as a distributed ledger, Ethereum is better characterized as a distributed state machine. The state of Ethereum is a large data structure that not only records all accounts and balances but also maintains a machine state that can transition between blocks according to a predefined set of rules and execute arbitrary machine code\cite{liu2024gastrace}. The specific rules governing state transitions within blocks are defined by the Ethereum Virtual Machine.\par

The Ethereum state contains a vast number of transactions stored within blocks, which are linked sequentially over time. Each time a new block is generated, it must be validated through a consensus algorithm to ensure network-wide agreement.\par

\subsubsection{Account}
The global state of Ethereum consists of individual accounts, each with its own state and a unique 20-byte address. Ethereum accounts are categorized into externally owned accounts and contract accounts\cite{li2024detecting}. EOAs, controlled by external private keys, are not associated with any code and are entirely managed by their respective private key holders. In contrast, contract accounts are governed exclusively by the smart contract code deployed on the Ethereum network. Both types of accounts can receive, hold, and transfer Ether and tokens, as well as interact with deployed smart contracts. An EOA can initiate a transaction by signing it with its private key to transfer assets to another EOA or a contract account. When a transaction is sent to a contract account, it triggers the execution of the associated contract code\cite{wang2023account, thakur2017authentication}. Unlike EOAs, contract accounts cannot independently initiate transactions; they can only generate transactions in response to received transactions through their triggered code execution.\par

\subsubsection{State}
In the context of Ethereum, the state is a large data structure known as the modified Merkle Patricia Trie, which links all accounts through hashes, allowing them to be traced back to a single root hash stored on the blockchain. The state is divided into account state and global state, and these two types of state will be described separately. The account state consists of four components: the nonce, balance, storage root, and code hash.\par

\begin{itemize}
\item {\texttt{Nonce}}: A block records transactions and states within a specific period, serving as the fundamental storage unit of the blockchain and a collection of transactions that have been completed.
\item {\texttt{Balance}}: A chain structure that links blocks in chronological order using hashes.
\item {\texttt{StorageRoot}}: All operations in the blockchain network are treated as transactions, also known as operations, through which records are generated on the blocks and the corresponding states are altered.
\item {\texttt{CodeHash}}: For externally owned accounts (EOAs), the corresponding field is empty. In contrast, for contract accounts, this field stores the hash of the account’s contract code.
\end{itemize}

The global state of Ethereum is represented as a mapping from account addresses to their corresponding account states. This mapping is maintained in a data structure known as the Merkle Patricia Trie (MPT), a specialized form of binary tree. The MPT is composed of a set of nodes and exhibits the following two properties.

\begin{itemize}
\item A large number of leaf nodes that contain the underlying data.
\item Each parent node stores the hash values of its two child nodes.

\end{itemize}

The MPT enables lightweight clients in the blockchain network to process information such as transactions, events, and balances without storing the entire blockchain. Due to the hash propagation property of the MPT, maliciously submitted falsified data (e.g., fake transactions) can be effectively prevented. By verifying the hashes in the block header, all nodes can validate a small subset of Ethereum’s global state without needing to maintain the full chain.\par

\subsubsection{Transaction Fees}
Similar to typical blockchain systems, all transactions on Ethereum require the payment of a transaction fee, known as gas. Users specify a gasLimit to cap the maximum amount of gas they are willing to consume for a transaction, and a gasPrice to determine the amount they are willing to pay per unit of gas. If the gas provided is insufficient to complete the transaction, the transaction fails, and all state changes made during its execution are reverted. Only when sufficient gas is supplied will the transaction be validated and confirmed. The consumed gas serves as a reward for miners (or, in the context of Ethereum 2.0, validators).\par

Miners have the autonomy to select which transactions they wish to validate or ignore. As a result, when constructing a block, miners tend to prioritize transactions with higher gas prices in order to maximize their rewards. Consequently, transaction initiators often increase the gas price to improve the likelihood of their transactions being included in a block.\par

For contract-related transactions, gas is also required to cover the additional costs of computation and storage\cite{honari2023smart, chen2021maintenance}. Computation on Ethereum is intentionally expensive, a design choice aimed at safeguarding the network’s integrity: every computation performed on-chain incurs a cost, thereby discouraging the submission of spam or malicious activity. To further mitigate risks such as unintended or malicious infinite loops and other forms of computational waste, each transaction must specify a limit on the number of computation steps it is allowed to execute.\par

\subsubsection{Transaction}

Transactions are the fundamental operations in Ethereum, enabling the transition of the system from one state to another. There are two types of transactions: message calls and contract creation. Both types share the same structure, which includes the following fields.\par

\begin{itemize}
\item {\texttt{Nonce}}: A sequential number indicating the sender's transaction count.
\item {\texttt{Gas Price}}: The amount the sender is willing to pay per unit of gas.
\item {\texttt{Gas Limit}}: The maximum amount of gas the sender is willing to provide for the transaction.
\item {\texttt{To (Recipient Address)}}: The address of the recipient. This field is left empty (i.e., set to the zero address) for contract creation transactions.
\item {\texttt{Value}}:The amount of Ether to be transferred to the recipient address.
\item {\texttt{Init}}:A field used only in contract creation transactions. It contains the initialization code for the smart contract. Upon execution, it returns the address of the newly deployed contract.
\item {\texttt{Data}}:A field used only in message call transactions. It carries input parameters to be passed during the invocation of the contract function.

\end{itemize}

\section{Implementation of Ethereum Token Contracts}
\subsection{Token Standard Interface}
Ethereum smart contracts offer powerful and reliable execution capabilities. Once predefined conditions are met, the contract automatically executes according to the logic encoded within it. This makes smart contracts particularly well-suited for applications in the domain of digital assets. One common use case is the development of tokens on the Ethereum platform. A token represents a unit of value or ownership on the blockchain. Like traditional cryptocurrencies, tokens can be transferred between contracts, queried for their total supply, or checked for individual account balances.\par

In late 2015, Fabian Vogelsteller proposed the ERC-20 standard (also known as EIP-20), which defines a standardized interface capturing these token characteristics\cite{dietrich2023blockchain}. This interface enables Ethereum wallets and other contracts to interact with tokens in a unified manner. ERC-20 specifies a set of rules that all fungible Ethereum tokens should follow, allowing developers to predict how new tokens will behave within the broader Ethereum ecosystem. For instance, token exchange protocols like Uniswap can support any newly issued ERC-20-compliant token without modification.\par

The ERC-20 standard is closely associated with Initial Coin Offering contracts, where a predefined amount of Ether is raised, and corresponding tokens are issued to users once the fundraising goal is met\cite{cuffe2018role}. ERC-20 defines only the standard interface and not its concrete implementation; developers must implement and maintain the actual contract code themselves. Since ERC-20 tokens focus solely on token quantity and treat all units as identical and interchangeable, they are classified as fungible tokens, and ERC-20 is the de facto standard for such tokens\cite{rahimian2021tokenhook, soni2023erc, viswasreddy2024efficient}.\par

In contrast, non-fungible tokens, where each token is unique and distinguishable from others, follow the ERC-721 standard\cite{kong2024characterizing, niu2024unveiling}. This type of token is widely used in scenarios involving digital collectibles, intellectual property rights, and digital agreements. The core interface includes the following functions:\par

\begin{itemize}
\item {\texttt{totalSupply}}: Returns the total supply of the token. Although token supply is usually fixed, this function allows the contract to return the effective circulating supply.

\item {\texttt{balanceOf}}: Takes an address as input and returns the token balance of that address. All token balances are publicly visible.

\item {\texttt{transferFrom}}: Transfers a specified number of tokens from one address to another. This operation must emit a Transfer event. It is primarily used to allow a designated spender to transfer tokens on behalf of the owner. The number of tokens that can be transferred is constrained by the allowance, which must be set via the approve function.

\item {\texttt{approve}}: Grants permission to a spender to withdraw a specified number of tokens from the owner’s account.

\end{itemize}

\subsection{Contract Compilation and Deployment}
\subsubsection{Tools}
 
\begin{itemize}
\item {\texttt{MetaMask}}: MetaMask is currently the most widely used Ethereum wallet and gateway. Available as a browser extension and mobile application, it manages digital assets through a secure local key vault. MetaMask enables users to buy, store, send, and exchange tokens, and also provides a simple and secure interface for connecting to Ethereum-based blockchain applications via RPC endpoints. This allows seamless interaction with decentralized applications directly from web pages using the local wallet.

\item {\texttt{TestNet}}: All nodes operating on the same blockchain are considered part of a single network. The network most commonly used in practice is the main network (MainNet), where users conduct real transactions and deploy production smart contracts. Due to its large user base, the MainNet provides high security and strong resistance against attacks and tampering, making it a robust and trustworthy environment. Consequently, Ether on the MainNet has real-world value, and testing on the MainNet incurs a significant cost. While a blockchain typically has only one MainNet, it may also support multiple test networks (TestNets) designed for experimentation, learning, and testing. A TestNet is an independent blockchain that starts from a different genesis block and may employ a different consensus mechanism than the MainNet. Users can mine blocks and perform testing without incurring real-world costs, making TestNets essential for safe and effective development workflows.

\end{itemize}

\subsubsection{Deployment}
After compiling the contract in Remix, the Injected Web3 environment is used to connect to the Rinkeby test network via MetaMask. The constructor parameter is set to 2048, indicating the issuance of a total of 2048 tokens. All tokens are initially assigned to the deployer's own address to facilitate testing.\par

After confirming the transaction and paying the gas fee, the deployment is completed once the transaction is mined and confirmed. The contract can then be inspected via Etherscan. The transaction was confirmed in block number 10593657 on the test network. The previously empty To field is filled with the generated contract address. Since the transaction is only intended to deploy or invoke a contract, the Ether transfer amount can be zero. Given that a sufficient amount of gas was provided, the transaction was successfully executed and all allocated gas was consumed.\par

As discussed earlier, Ethereum operates as a distributed state machine. This transaction resulted in state transitions for three different account addresses. As shown in Figures 4 and 5, a new contract address was created with associated storage space, and the sender’s account balance was reduced due to the gas fee. Additionally, the nonces of both the sender and the contract addresses were incremented by one after the transaction.\par

\begin{figure*}[htbp]
     \centering
    \includegraphics[height=4cm,width=16cm]{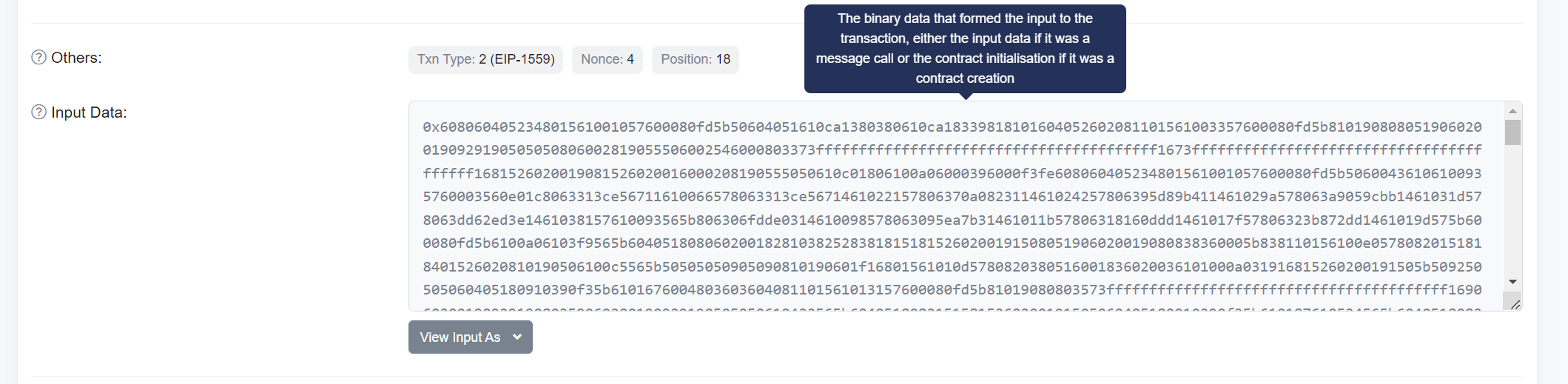}
    \caption{Contract Bytecode Embedded in the Transaction.}
    \label{fig:4}
\end{figure*}

\begin{figure*}[htbp]
     \centering
    \includegraphics[height=4cm,width=16cm]{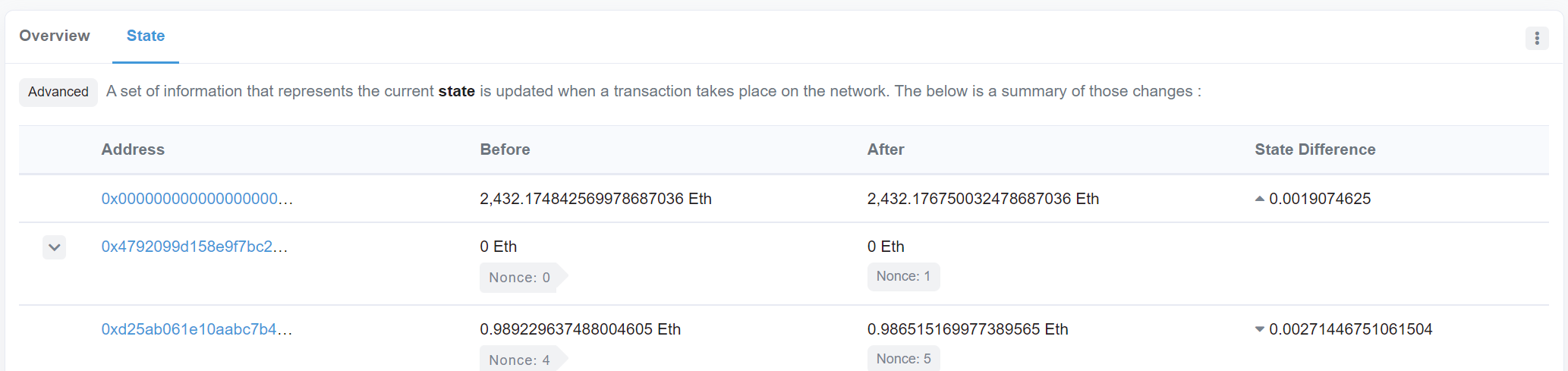}
    \caption{State Transitions Caused by the Transaction.}
    \label{fig:5}
\end{figure*}

As illustrated in Figure 6, the contract's storage was initially empty upon deployment. This transaction initializes the storage layout. Specifically, the first storage slot holds the totalSupply\_ variable set during construction, while the second slot stores the mapping entry that associates the current wallet address with its token balance.\par

\begin{figure*}[htbp]
     \centering
    \includegraphics[height=4cm,width=16cm]{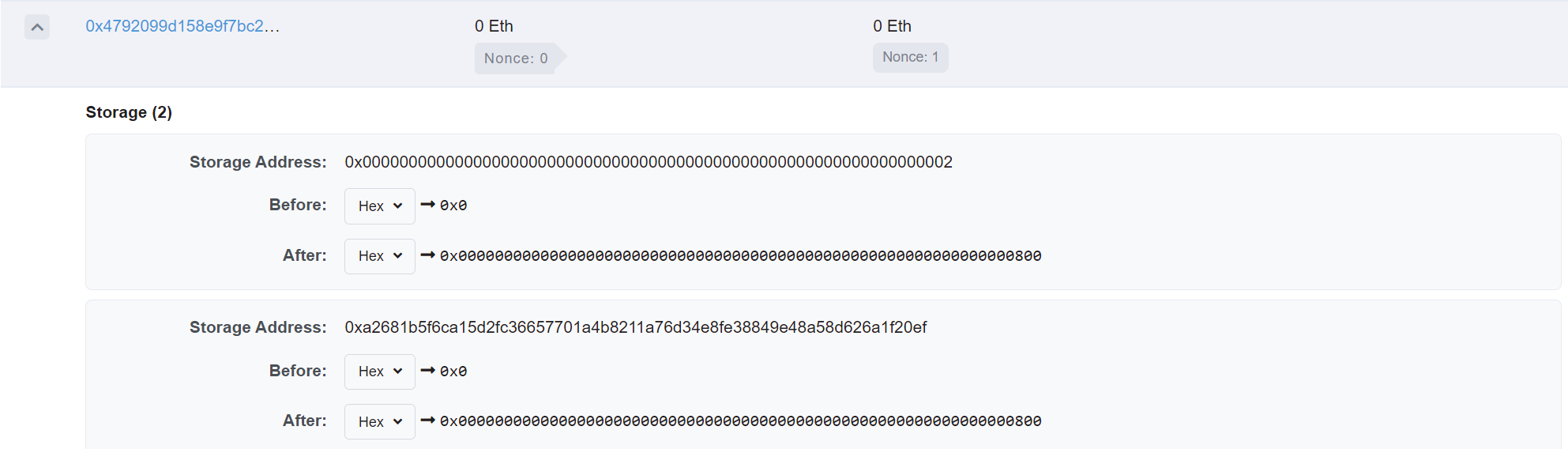}
    \caption{Storage Changes at the Contract Address Resulting from the Transaction.}
    \label{fig:6}
\end{figure*}

As shown in Figure 7, the transaction that invoked the contract included a sequence of encoded data. This data encodes the function selector followed by its parameters in order: the target address (parameter 1) and the amount to transfer (parameter 2).\par

\begin{figure*}[htbp]
     \centering
    \includegraphics[height=11cm,width=13.5cm]{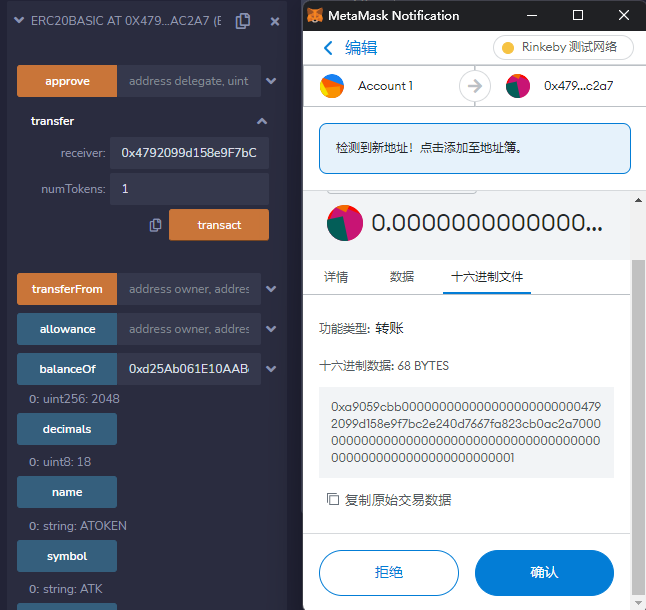}
    \caption{Token Transfer Invocation via Remix’s Contract Interaction Interface.}
    \label{fig:7}
\end{figure*}

Due to the standardized ERC-20 interface, Etherscan can automatically recognize the token properties and generate a token tracking page with key metadata, as shown in Figure 8. As a result of the transfer operation performed in this transaction, the number of token-holding addresses increased from one to two, with token balances of 2047 and 1, respectively.\par

\begin{figure*}[htbp]
     \centering
    \includegraphics[height=8cm,width=16cm]{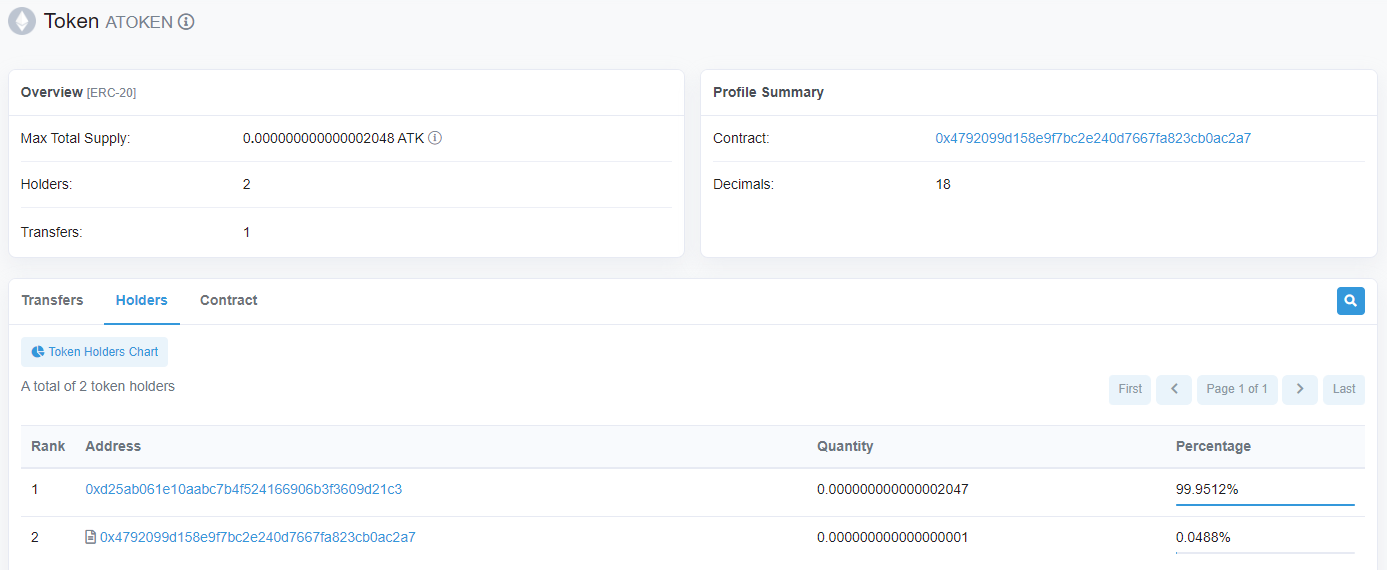}
    \caption{Token Tracking Page for ATOKEN on Etherscan.}
    \label{fig:8}
\end{figure*}

\section{Common Security Threats in Token Contracts}
Smart contract vulnerabilities refer to potential security flaws in the contract code that, if exploited by attackers, can result in asset losses within the contract\cite{li2024stateguard, li2025scalm}. Since The DAO attack, hackers have realized that smart contracts represent a lucrative target, leading to an era of active vulnerability hunting. Numerous security issues have since been uncovered. In this paper, we classify these vulnerabilities into two categories: those arising from the design characteristics of the Ethereum platform, and those caused by traditional attack techniques. Classic vulnerabilities are reproduced and analyzed to better understand their causes and consequences.\par
\subsection{Ethereum Platform-Induced Vulnerabilities}
\subsubsection{Re-Entrancy Vulnerability}
Re-entrancy is a vulnerability where an attacker exploits the fallback function to recursively invoke a vulnerable transfer method, allowing repeated withdrawals before the contract's state is correctly updated. This recursive behavior continues until the transaction either runs out of gas or reaches a specified termination condition. As a result, attackers can withdraw tokens far exceeding their initial deposit.\par

The fallback function is a special unnamed function in a contract, and each contract may define only one such function. In versions prior to Solidity 0.4.x, visibility was not strictly required. From version 0.5.x onward, fallback functions must be explicitly marked as external. These functions can be virtual, overridden, or decorated with modifiers, and are triggered in the following two scenarios:

\begin{itemize}
\item {\texttt{Function Not Found}}: If a function call does not match any defined function signatures in the contract, the fallback function is automatically invoked. This means that any invalid call (i.e., one with an undefined function selector) will result in fallback execution.

\item {\texttt{Using send() to Transfer Ether}}: When Ether is sent to a contract via the send() method without any attached data, the fallback function is invoked. As long as no valid contract method is explicitly called, the fallback function will be executed by default.
\end{itemize}

The infamous The DAO vulnerability on Ethereum falls into this category. This vulnerability led to over \$10 million in losses and directly triggered the Ethereum hard fork. It also inspired a wave of vulnerability research in the smart contract community. In the payout method, if the \_recipient is a contract address, the call instruction will invoke the recipient's fallback function. Since call does not restrict gas usage, the fallback function can recursively invoke payout, leading to a re-entrant attack that drains the contract's balance. After deploying the vulnerable contract, we fund it by transferring Ether to its address. Based on the vulnerability, we can write an attack contract as shown in Listing 1. The core part of the attack is implemented in the fallback function, where a call to the splitDAO function of the TheDAO contract is made to achieve the reentrancy effect. A limit on the number of iterations is set to prevent the transaction from reverting due to gas exhaustion. Finally, a withdrawal function is used to extract the Ether obtained from the attack.\par

\begin{listing}[H]%
\caption{Attack contract}%
\label{lst:listing1}%
\begin{lstlisting}
contract HackCode {
    address public daoContract;
    uint public count = 50;
    uint public n;
    function setDAO(address _addr) public {
        daoContract = _addr;
    }
    function getBalance() public view returns (uint) {
        return address(this).balance;
    }
    function withdraw() public {
        msg.sender.transfer(address(this).balance);
    }
    function setCount(uint newCount) public {
        count = newCount;
    }
    function () public payable {
        if(n < count){
            n++;
            TheDAO(daoContract).splitDAO();
        }
    }
}
\end{lstlisting}
\end{listing}

 Attack Execution Process:
\begin{itemize}
\item Deploy the attack contract.
\item Invoke the setDAO method via a transaction to set the target address variable to the victim contract.
\item Invoke the setCount method via a transaction to set the count to an appropriate value.
\item Trigger the fallback function of the attack contract through a regular transaction.
\item The attack contract will call the victim DAO contract’s splitDAO method, which internally invokes withdrawRewardFor, and ultimately calls payout. Within payout, a call is made to transfer Ether to the attack contract, which triggers the fallback function again, leading to re-entrancy. The splitDAO method is entered again, causing Ether to be withdrawn repeatedly. Even if a check is placed after the transfer, it will not take effect during this process.
\item Upon a successful attack, the attack contract extracts most of the Ether from the victim contract. At this point, the attacker can call withdraw to retrieve the stolen funds from the attack contract and end the attack.

\end{itemize}

Mitigation and Prevention:

\begin{itemize}
\item Whenever possible, use send or transfer instead of call. The gasLimit of send and transfer is 2300, which is insufficient to support even the simplest function call, thus preventing fallback execution.
\item For functions involving transfers and payments, adopt a "checks-effects-interactions" pattern: first validate conditions, then update state, and finally perform the transfer. This helps prevent many potential issues.

\end{itemize}

\subsubsection{Delegatecall Vulnerability}
In Solidity, there are two methods to invoke external contracts: call and delegatecall, each with different contextual behaviors.\par

\begin{itemize}
\item {\texttt{Call}}: When using call to invoke an external contract method, the execution context is that of the external contract. When Contract A calls a function of external Contract B via call, it executes in the context of Contract B and then returns to Contract A to continue execution.

\item {\texttt{delegatecall}}: When using delegatecall to invoke an external contract method, the execution context is that of the local contract. When Contract A calls a function of external Contract B via delegatecall, it is equivalent to copying Contract B’s code and executing it within Contract A’s context. 
\end{itemize}

Attack Execution Process:

\begin{itemize}
\item  Account 1 deploys the Delegate contract.

\item Account 1 deploys the Delegation contract, specifying the address of the Delegate contract to act as its proxy.

\item At this point, verify that the owner of both contracts is the same, i.e., the address of Account 1.
\item Account 2 invokes the fallback function of Delegation, modifying the owner address.
\item Check the owner of Delegation, which has now been changed to the address of Account 2.

\end{itemize}

Mitigation and Prevention:

\begin{itemize}
\item Use delegatecall with caution, and clearly define function visibility. Sensitive functions should be declared as external to prevent unintended external invocation.
\end{itemize}

\subsection{Traditional Vulnerabilities Reproduced on the Ethereum Platform}

\subsubsection{Integer Overflow Vulnerability}

Due to the limited number of bits that a register can represent, when a stored value exceeds the maximum representable range, overflow may occur. Maximum value overflow wraps around to the minimum value, and minimum value overflow wraps around to the maximum. This issue can occur on any platform and is one of the most common and universal types of vulnerabilities [25]. Similarly, the Ethereum Virtual Machine (EVM) assigns fixed-size data types for integers. An integer variable can only represent values within a specific range. For example, the largest integer type is uint256, which has a maximum value of 2**256 - 1. Exceeding this range will result in overflow.\par

 In the transfer condition check, require(balances[msg.sender] - \_value >= 0) presents a clear integer overflow issue. Since uint is actually uint256, transferring more than the initial balance causes an underflow, turning (balances[msg.sender] - \_value) into a large positive integer. This allows transferring more tokens than the actual balance and inflates the sender’s balance to an extremely large number. The attack method involves directly calling the transfer function with a value greater than the available balance.\par

To protect contracts from overflow vulnerabilities, arithmetic functions in the SafeMath library are typically used to replace regular addition, subtraction, multiplication, and division. However, any oversight may lead to severe vulnerabilities.\par

As shown in Listing 3, the vulnerability that once caused BeautyChain tokens to be minted infinitely and their value to drop to zero is of this integer overflow type. Although the contract included and used the SafeMath library, one instance of default multiplication led to the problem.\par

\begin{listing}[H]%
\caption{The vulnerability function of Meitu Coin}%
\label{lst:listing2}%
\begin{lstlisting}
contract BeautyChain {
    using SafeMath for uint256;
    mapping (address => uint256) public balances;

    function batchTransfer(address[] _receivers, uint256 _value) public returns (bool) {
        uint cnt = _receivers.length;
        uint256 amount = uint256(cnt) * _value;
        require(cnt > 0 && cnt <= 20);
        require(_value > 0 && balances[msg.sender] >= amount);

        balances[msg.sender] = balances[msg.sender].sub(amount);
        for (uint i = 0; i < cnt; i++) {
            balances[_receivers[i]] = balances[_receivers[i]].add(_value);
        }

        return true;
    }
}

\end{lstlisting}
\end{listing}

The variable amount is then used as a condition in a subsequent transfer operation. Since this is a public function, both \_receivers and \_value are controllable. By causing amount to overflow upwards into a very small value, the validation can be bypassed to allow the transfer of a large value. A feasible set of attack vectors can be constructed accordingly.\par

\begin{itemize}
\item Set \_receivers as an array containing two recipient addresses, such that \_receivers.length = 2.
\item Set \_value = 0x8000000000000000000000000000000000000000000000000000000000000000.
\end{itemize}

Under this condition, \_receivers.length * \_value causes an overflow, resulting in amount = 0, thereby bypassing the balance check and allowing the transfer of the aforementioned amount of tokens.\par

Fix and Prevention:

\begin{itemize}
\item Simply replacing * with the mul function from the SafeMath library can prevent the overflow issue.
\end{itemize}

Such problems are easy to detect through tooling and auditing. When performing arithmetic operations in a contract, overflow checks must be properly implemented. Starting from Solidity version 0.8.0, all arithmetic operations include built-in overflow checks by default, eliminating the need for external libraries. Therefore, using a newer version of Solidity also contributes to improved security.\par

\subsubsection{Random Number Predictability}

Random numbers are widely used in scenarios such as lotteries, games, and signature algorithms. They form the foundation of cryptography and privacy security in traditional internet applications. However, achieving true randomness on the blockchain is particularly challenging. As a distributed consensus network, everything on the blockchain is public—including algorithms, variables, and states—and there is often no suitable source of entropy, making random numbers predictable. Platforms often rely on on-chain public information as randomness sources. Each platform has unique attributes that are publicly accessible; therefore, using such attributes as seeds for randomness undermines unpredictability. Many real-world attacks have occurred on platforms like Ethereum and EOS.\par

In this example, the most significant bit of the previous blockhash is used as the random bit for the coin flip. Since the hash of the previous block is fully known, one can observe it and then call the guessing function before the next block is mined. An attacker can deploy a contract to predict and interact with the vulnerable contract, as shown in Listing 3, achieving consecutive wins through random number prediction.\par

\begin{listing}[H]%
\caption{Prediction attack contract targeting the coin-flip contract}%
\label{lst:listing2}%
\begin{lstlisting}
contract Attack {
    CoinFlip fliphack;
    address victim;
    uint256 FACTOR = 57896044618658097711785492504343953926634992332820282019728792003956564819968;

    function Attack(address victim) {
        fliphack = CoinFlip(victim);
    }
    function predict() public view returns (bool) {
        uint256 blockValue = uint256(block.blockhash(block.number - 1));
        uint256 coinFlip = uint256(uint256(blockValue) / FACTOR);
        return coinFlip == 1 ? true : false;
    }
    function hack() public {
        bool guess = predict();
        fliphack.flip(guess);
    }
}

\end{lstlisting}
\end{listing}
 
In 2018, the FoMo3D contract on Ethereum suffered a prediction attack on its random airdrop algorithm because it used the previous block hash as the entropy source. The attacker used random number prediction to decide in advance whether to participate, thereby capturing large airdrop rewards.\par

Mitigation Strategies:

\begin{itemize}
\item Since all content on the blockchain is transparent to participants, using randomness in Ethereum is inherently complex. However, several approaches can improve security. For instance, using less predictable pseudo-random sources such as block timestamps, or relying on off-chain oracles to generate randomness for on-chain use [27].
\end{itemize}

\section{Conclusion}
This paper uses Ethereum as a case study to analyze the technical applications of cryptocurrency and associated security issues in Blockchain 2.0, characterized by support for smart contracts. Compared to traditional application platforms, smart contracts represent a relatively new paradigm. While they generally consist of smaller codebases, increasing complexity combined with insufficiently tested code and inherent platform features still present many security vulnerabilities. For a blockchain network that supports millions of contracts and a value-based ecosystem, security is a highly sensitive concern. Despite the current low levels of security assurance, substantial investment in blockchain will continue to drive security research. As new attack patterns and vulnerabilities are discovered, classification of related issues will continue to evolve.Research into smart contract security contributes to enhancing their robustness and supports the development of a more secure and reliable Blockchain 2.0 environment. Currently, the primary method for ensuring contract security involves auditing and detection-based validation [28], along with formal verification of certain components [29]. However, existing tools and audits cannot eliminate security risks. Many contracts that have passed both automated and manual checks still exhibit vulnerabilities. Thus, the security of smart contracts needs further reinforcement, and the development of corresponding security detection tools remains an open and critical area for advancement.\par

\bibliographystyle{ACM-Reference-Format}
\bibliography{main}

@String{Computing = "Computing" }

@String{Computer = "{IEEE} Computer" }

@String{Springer = "Springer-Verlag" }

@incollection{mukherjee2021blockchain,
  title={Blockchain 1.0 to blockchain 4.0—The evolutionary transformation of blockchain technology},
  author={Mukherjee, Pratyusa and Pradhan, Chittaranjan},
  booktitle={Blockchain technology: applications and challenges},
  pages={29--49},
  year={2021},
  publisher={Springer}
}

@incollection{aggarwal2021blockchain,
  title={Blockchain 2.0: smart contracts},
  author={Aggarwal, Shubhani and Kumar, Neeraj},
  booktitle={Advances in computers},
  volume={121},
  pages={301--322},
  year={2021},
  publisher={Elsevier}
}

@article{rahimian2021tokenhook,
  title={TokenHook: Secure ERC-20 smart contract},
  author={Rahimian, Reza and Clark, Jeremy},
  journal={arXiv preprint arXiv:2107.02997},
  year={2021}
}

@article{soni2023erc,
  title={ERC-20 Token Exchange System Over Blockchain Network},
  author={Soni, Mukund and Gandotra, Ekta},
  year={2023},
  publisher={Jaypee University of Information Technology, Solan, HP}
}

@inproceedings{viswasreddy2024efficient,
  title={Efficient Token Transfer Using ERC-20 Decentralized Exchange},
  author={ViswasReddy, Mettupalle Chinnaiahgari Venkata and Reddy, Vootkuri Sai Charan and Rao, Gudeme Jaya and Devi, N Rama},
  booktitle={2024 International Conference on Intelligent Systems for Cybersecurity (ISCS)},
  pages={01--06},
  year={2024},
  organization={IEEE}
}

@article{honari2023smart,
  title={Smart contract design in distributed energy systems: a systematic review},
  author={Honari, Kimia and Rouhani, Sara and Falak, Nida E and Liu, Yuan and Li, Yunwei and Liang, Hao and Dick, Scott and Miller, James},
  journal={Energies},
  volume={16},
  number={12},
  pages={4797},
  year={2023},
  publisher={MDPI}
}

@article{chen2021maintenance,
  title={Maintenance-related concerns for post-deployed Ethereum smart contract development: issues, techniques, and future challenges},
  author={Chen, Jiachi and Xia, Xin and Lo, David and Grundy, John and Yang, Xiaohu},
  journal={Empirical Software Engineering},
  volume={26},
  number={6},
  pages={117},
  year={2021},
  publisher={Springer}
}

@article{yu2020blockchain,
  title={Blockchain-based shared manufacturing in support of cyber physical systems: concept, framework, and operation},
  author={Yu, Chunyang and Jiang, Xuanlin and Yu, Shiqiang and Yang, Cheng},
  journal={Robotics and Computer-Integrated Manufacturing},
  volume={64},
  pages={101931},
  year={2020},
  publisher={Elsevier}
}

@article{sedlmeir2020energy,
  title={The energy consumption of blockchain technology: Beyond myth},
  author={Sedlmeir, Johannes and Buhl, Hans Ulrich and Fridgen, Gilbert and Keller, Robert},
  journal={Business \& Information Systems Engineering},
  volume={62},
  number={6},
  pages={599--608},
  year={2020},
  publisher={Springer}
}

@techreport{budish2018economic,
  title={The economic limits of bitcoin and the blockchain},
  author={Budish, Eric},
  year={2018},
  institution={National Bureau of Economic Research}
}

@article{andoni2019blockchain,
  title={Blockchain technology in the energy sector: A systematic review of challenges and opportunities},
  author={Andoni, Merlinda and Robu, Valentin and Flynn, David and Abram, Simone and Geach, Dale and Jenkins, David and McCallum, Peter and Peacock, Andrew},
  journal={Renewable and sustainable energy reviews},
  volume={100},
  pages={143--174},
  year={2019},
  publisher={Elsevier}
}

@inproceedings{kong2024characterizing,
  title={Characterizing the Solana NFT ecosystem},
  author={Kong, Dechao and Li, Xiaoqi and Li, Wenkai},
  booktitle={Companion Proceedings of the ACM Web Conference 2024},
  pages={766--769},
  year={2024}
}

@inproceedings{li2024cobra,
  title={Cobra: interaction-aware bytecode-level vulnerability detector for smart contracts},
  author={Li, Wenkai and Li, Xiaoqi and Li, Zongwei and Zhang, Yuqing},
  booktitle={Proceedings of the 39th IEEE/ACM International Conference on Automated Software Engineering},
  pages={1358--1369},
  year={2024}
}

@article{li2024scla,
  title={SCLA: Automated Smart Contract Summarization via LLMs and Semantic Augmentation},
  author={Mao, Yingjie and Li, Xiaoqi and Li, Wenkai and Wang, Xin and Xie, Lei},
  journal={arXiv preprint arXiv:2402.04863},
  year={2024}
}

@inproceedings{li2024detecting,
  title={Detecting Malicious Accounts in Web3 through Transaction Graph},
  author={Li, Wenkai and Liu, Zhijie and Li, Xiaoqi and Nie, Sen},
  booktitle={Proceedings of the 39th IEEE/ACM International Conference on Automated Software Engineering},
  pages={2482--2483},
  year={2024}
}

@article{albert2024superstack,
  title={SuperStack: Superoptimization of stack-bytecode via greedy, constraint-based, and SAT techniques},
  author={Albert, Elvira and Garcia de la Banda, Maria and Hern{\'a}ndez-Cerezo, Alejandro and Ignatiev, Alexey and Rubio, Albert and Stuckey, Peter J},
  journal={Proceedings of the ACM on Programming Languages},
  volume={8},
  number={PLDI},
  pages={1437--1462},
  year={2024},
  publisher={ACM New York, NY, USA}
}

@article{hildenbrandt2017kevm,
  title={Kevm: A complete semantics of the ethereum virtual machine},
  author={Hildenbrandt, Everett and Saxena, Manasvi and Zhu, Xiaoran and Rodrigues, Nishant and Daian, Philip and Guth, Dwight and Ro{\c{s}}u, Grigore},
  year={2017}
}

@article{fekete2023toward,
  title={Toward building smart contract-based higher education systems using zero-knowledge Ethereum virtual machine},
  author={Fekete, D{\'e}nes L{\'a}szl{\'o} and Kiss, Attila},
  journal={Electronics},
  volume={12},
  number={3},
  pages={664},
  year={2023},
  publisher={MDPI}
}

@article{wang2019survey,
  title={A survey on consensus mechanisms and mining strategy management in blockchain networks},
  author={Wang, Wenbo and Hoang, Dinh Thai and Hu, Peizhao and Xiong, Zehui and Niyato, Dusit and Wang, Ping and Wen, Yonggang and Kim, Dong In},
  journal={Ieee Access},
  volume={7},
  pages={22328--22370},
  year={2019},
  publisher={IEEE}
}

@inproceedings{li2024stateguard,
  title={StateGuard: Detecting State Derailment Defects in Decentralized Exchange Smart Contract},
  author={Li, Zongwei and Li, Wenkai and Li, Xiaoqi and Zhang, Yuqing},
  booktitle={Companion Proceedings of the ACM Web Conference 2024},
  pages={810--813},
  year={2024}
}

@article{li2025scalm,
  title={SCALM: Detecting Bad Practices in Smart Contracts Through LLMs},
  author={Li, Zongwei and Li, Xiaoqi and Li, Wenkai and Wang, Xin},
  journal={arXiv preprint arXiv:2502.04347},
  year={2025}
}

@inproceedings{liu2024gastrace,
  title={GasTrace: Detecting Sandwich Attack Malicious Accounts in Ethereum},
  author={Liu, Zekai and Li, Xiaoqi and Peng, Hongli and Li, Wenkai},
  booktitle={2024 IEEE International Conference on Web Services (ICWS)},
  pages={1409--1411},
  year={2024},
  organization={IEEE}
}

@inproceedings{niu2024unveiling,
  title={Unveiling wash trading in popular NFT markets},
  author={Niu, Yuanzheng and Li, Xiaoqi and Peng, Hongli and Li, Wenkai},
  booktitle={Companion Proceedings of the ACM Web Conference 2024},
  pages={730--733},
  year={2024}
}

@article{curry2025limitations,
  title={Limitations of trust and legitimacy in blockchain: exploring the effectiveness of decentralisation, immutability and consensus mechanisms in blockchain governance},
  author={Curry, Dion},
  journal={International Journal of Public Sector Management},
  volume={38},
  number={1},
  pages={98--117},
  year={2025},
  publisher={Emerald Publishing Limited}
}

@article{ma2025understanding,
  title={Understanding Security Issues in the DAO Governance Process},
  author={Ma, Junjie and Jiang, Muhui and Jiang, Jinan and Luo, Xiapu and Hu, Yufeng and Zhou, Yajin and Wang, Qi and Zhang, Fengwei},
  journal={IEEE Transactions on Software Engineering},
  year={2025},
  publisher={IEEE}
}

@inproceedings{dika2018security,
  title={Security vulnerabilities in ethereum smart contracts},
  author={Dika, Ardit and Nowostawski, Mariusz},
  booktitle={2018 IEEE international conference on Internet of Things (iThings) and IEEE green computing and communications (GreenCom) and IEEE cyber, physical and social computing (CPSCom) and IEEE Smart Data (SmartData)},
  pages={955--962},
  year={2018},
  organization={IEEE}
}

@article{kushwaha2022ethereum,
  title={Ethereum smart contract analysis tools: A systematic review},
  author={Kushwaha, Satpal Singh and Joshi, Sandeep and Singh, Dilbag and Kaur, Manjit and Lee, Heung-No},
  journal={Ieee Access},
  volume={10},
  pages={57037--57062},
  year={2022},
  publisher={IEEE}
}

@article{duy2025vulnsense,
  title={Vulnsense: Efficient vulnerability detection in ethereum smart contracts by multimodal learning with graph neural network and language model},
  author={Duy, Phan The and Khoa, Nghi Hoang and Quyen, Nguyen Huu and Trinh, Le Cong and Kien, Vu Trung and Hoang, Trinh Minh and Pham, Van-Hau},
  journal={International Journal of Information Security},
  volume={24},
  number={1},
  pages={48},
  year={2025},
  publisher={Springer}
}

@article{liang2025vulseye,
  title={Vulseye: Detect smart contract vulnerabilities via stateful directed graybox fuzzing},
  author={Liang, Ruichao and Chen, Jing and Wu, Cong and He, Kun and Wu, Yueming and Cao, Ruochen and Du, Ruiying and Zhao, Ziming and Liu, Yang},
  journal={IEEE Transactions on Information Forensics and Security},
  year={2025},
  publisher={IEEE}
}

@article{liu2022blockchain,
  title={A blockchain-based decentralized, fair and authenticated information sharing scheme in zero trust internet-of-things},
  author={Liu, Yizhi and Hao, Xiaohan and Ren, Wei and Xiong, Ruoting and Zhu, Tianqing and Choo, Kim-Kwang Raymond and Min, Geyong},
  journal={IEEE Transactions on Computers},
  volume={72},
  number={2},
  pages={501--512},
  year={2022},
  publisher={IEEE}
}

@article{al2022hierarchical,
  title={Hierarchical blockchain structure for node authentication in IoT networks},
  author={Al Ahmed, Mahmoud Tayseer and Hashim, Fazirulhisyam and Hashim, Shaiful Jahari and Abdullah, Azizol},
  journal={Egyptian Informatics Journal},
  volume={23},
  number={2},
  pages={345--361},
  year={2022},
  publisher={Elsevier}
}

@inproceedings{zhang2020overview,
  title={Overview of blockchain consensus mechanism},
  author={Zhang, Changqiang and Wu, Cangshuai and Wang, Xinyi},
  booktitle={Proceedings of the 2020 2nd International Conference on Big Data Engineering},
  pages={7--12},
  year={2020}
}

@article{wang2019blockchain,
  title={Blockchain-enabled smart contracts: architecture, applications, and future trends},
  author={Wang, Shuai and Ouyang, Liwei and Yuan, Yong and Ni, Xiaochun and Han, Xuan and Wang, Fei-Yue},
  journal={IEEE Transactions on Systems, Man, and Cybernetics: Systems},
  volume={49},
  number={11},
  pages={2266--2277},
  year={2019},
  publisher={IEEE}
}

@article{ferdous2021survey,
  title={A survey of consensus algorithms in public blockchain systems for crypto-currencies},
  author={Ferdous, Md Sadek and Chowdhury, Mohammad Jabed Morshed and Hoque, Mohammad A},
  journal={Journal of Network and Computer Applications},
  volume={182},
  pages={103035},
  year={2021},
  publisher={Elsevier}
}

@inproceedings{hirai2017defining,
  title={Defining the ethereum virtual machine for interactive theorem provers},
  author={Hirai, Yoichi},
  booktitle={Financial Cryptography and Data Security: FC 2017 International Workshops, WAHC, BITCOIN, VOTING, WTSC, and TA, Sliema, Malta, April 7, 2017, Revised Selected Papers 21},
  pages={520--535},
  year={2017},
  organization={Springer}
}

@article{mazumdar2019survey,
  title={A survey on data storage and placement methodologies for cloud-big data ecosystem},
  author={Mazumdar, Somnath and Seybold, Daniel and Kritikos, Kyriakos and Verginadis, Yiannis},
  journal={Journal of Big Data},
  volume={6},
  number={1},
  pages={1--37},
  year={2019},
  publisher={Springer}
}

@article{khan2021blockchain,
  title={Blockchain smart contracts: Applications, challenges, and future trends},
  author={Khan, Shafaq Naheed and Loukil, Faiza and Ghedira-Guegan, Chirine and Benkhelifa, Elhadj and Bani-Hani, Anoud},
  journal={Peer-to-peer Networking and Applications},
  volume={14},
  pages={2901--2925},
  year={2021},
  publisher={Springer}
}

@article{christidis2016blockchains,
  title={Blockchains and smart contracts for the internet of things},
  author={Christidis, Konstantinos and Devetsikiotis, Michael},
  journal={IEEE access},
  volume={4},
  pages={2292--2303},
  year={2016},
  publisher={Ieee}
}

@inproceedings{wang2023account,
  title={Account abstraction, analysed},
  author={Wang, Qin and Chen, Shiping},
  booktitle={2023 IEEE International Conference on Blockchain (Blockchain)},
  pages={323--331},
  year={2023},
  organization={IEEE}
}

@article{thakur2017authentication,
  title={Authentication, authorization and accounting with Ethereum blockchain},
  author={Thakur, Mukesh and others},
  journal={Helsingfors universitet},
  year={2017}
}

@article{dietrich2023blockchain,
  title={Blockchain-based traceability architecture for mapping object-related supply chain events},
  author={Dietrich, Fabian and Louw, Louis and Palm, Daniel},
  journal={Sensors},
  volume={23},
  number={3},
  pages={1410},
  year={2023},
  publisher={MDPI}
}

@article{cuffe2018role,
  title={The role of the erc-20 token standard in a financial revolution: the case of initial coin offerings},
  author={Cuffe, Paul},
  year={2018},
  publisher={IEC-IEEE-KATS}
}

\end{document}